\documentclass[preprint, showkeys, amsmath, amssymb]{revtex4}

\usepackage{graphicx}

\begin{document}

\title{The Influence of Horizontal Gene Transfer on the Mean Fitness of Unicellular Populations in Static Environments}

\author{Yoav Raz}
\affiliation{Department of Chemistry, Ben-Gurion University of the Negev, Be'er-Sheva, Israel}
\author{Emmanuel Tannenbaum}
\email{emanuelt@bgu.ac.il}
\affiliation{Department of Chemistry, Ben-Gurion University of the Negev, Be'er-Sheva, Israel}

\begin{abstract}

This paper develops a mathematical model describing the influence that conjugation-mediated Horizontal Gene Transfer (HGT) has on the mutation-selection balance in an asexually reproducing population of unicellular, prokaryotic organisms.  It is assumed that mutation-selection balance is reached in the presence of a fixed background concentration of antibiotic, to which the population must become resistant in order to survive.  We analyze the behavior of the model in the limit of low and high antibiotic-induced first-order death rate constants, and find that the highest mean fitness is obtained at low rates of bacterial conjugation.  As the rate of conjugation crosses a threshold, the mean fitness decreases to a minimum, and then rises asymptotically to a limiting value as the rate of conjugation becomes infinitely large.  However, this limiting value is smaller than the mean fitness obtained in the limit of low conjugation rate.  This dependence of the mean fitness on the conjugation rate is fairly small for the parameter ranges we have considered, and disappears as the first-order death rate constant due to the presence of antibiotic approaches zero.  For large values of the antibiotic death rate constant, we have obtained an analytical solution for the behavior of the mean fitness that agrees well with the results of simulations.  The results of this paper suggest that conjugation-mediated HGT has a slightly deleterious effect on the mean fitness of a population at mutation-selection balance.  Therefore, we argue that HGT confers a selective advantage by allowing for faster adaptation to a new or changing environment.  The results of this paper are consistent with the observation that HGT can be promoted by environmental stresses on a population.

\end{abstract}


\maketitle

\section{Introduction}

Horizontal Gene Transfer (HGT) is considered to be any form of direct transfer of genetic material between two organisms, where one organism is not the parent of the other (the latter case is known as {\it vertical gene transfer}) (Ochman et al. 2000).  HGT has become a subject of great interest for both molecular and evolutionary biologists, because it is believed that HGT plays a large role in re-shaping prokaryotic genomes (Ochman et al. 2000).  In particular, HGT is believed to be primarily responsible for the rapid spread of antibiotic drug resistance in bacterial populations (Walsh 2000).  Given that the emergence of antibiotic drug resistant strains of bacteria has become a major public health concern (Walsh 2000), an understanding of HGT is not only important for advancing our knowledge of biology, but it is also of immense practical significance.  

Currently, there are three known mechanisms by which HGT occurs (Ochman et al. 2000):
\begin{enumerate}
\item {\it Transformation}:  When an organism (generally a bacterium) collects genetic material from its environment.
\item {\it Transduction}:  When a virus directly infiltrates a bacterium with genetic material.
\item {\it Bacterial Conjugation}:  When a bacterium transfers genetic information via intercellular contact with another bacterium. 
\end{enumerate} 

Bacterial conjugation is believed to be the most important mechanism responsible for HGT (Ochman et al. 2000), and so, in this paper, we will focus on developing mathematical models describing the role that conjugation-mediated HGT has on the mutation-selection balance of bacterial populations.  Given the presumed importance that HGT has for the spread of antibiotic drug resistance in bacterial populations, the mathematical models we develop will look at the influence of HGT on the mutation-selection balance in the presence of an antibiotic.  

The best characterized bacterial conjugation system is the F$^+$/F$^-$ system (Russi et al. 2008).  Here, a bacterium containing what is termed an {\it F-plasmid} fuses with a bacterium lacking the F-plasmid.  The bacterium containing the F-plasmid is termed an F$^+$ bacterium while the bacterium that does not contain this plasmid is termed an F$^-$ bacterium.  When the F$^+$ bacterium meets an F$^-$ bacterium, it transfers one of the strands of the F-plasmid to the F$^-$ bacterium via a pilus.  Once a strand of the F-plasmid has been transferred from the F$^+$ bacterium to the F$^-$ bacterium, a copy of the plasmid in both cells is produced by daughter strand synthesis using the DNA template strands.  The F$^-$ bacterium then becomes an F$^+$ bacterium that transcribes its own pilus and is able to transfer the F$^+$ plasmid to other bacteria in the population (Russi et al. 2008).  This process is illustrated in Figure 1.

\begin{figure}
\includegraphics[width = 0.9\linewidth, angle = -90]{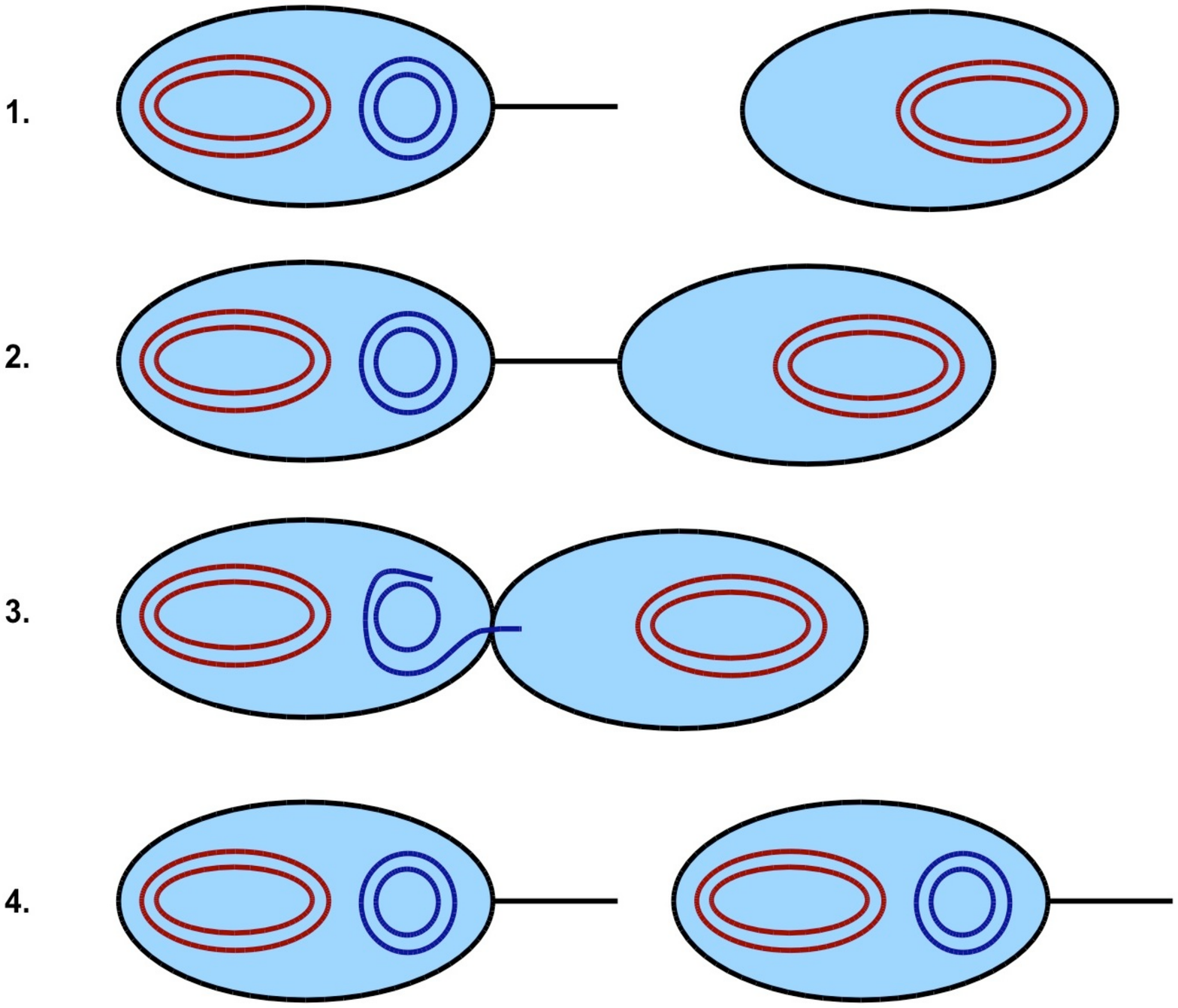}
\caption{Illustration of the process of bacterial conjugation.  In steps 1 and 2, an F$^+$ bacterium containing the F-plasmid (blue) binds to an F$^-$ bacterium lacking the plasmid.  One of the template strands from the F-plasmid then moves into the F$^-$ bacterium, as shown in step 3.  In step 4, the complementary strands are synthesized to reform the complete F-plasmids in both bacteria.  Both bacteria are now of the F$^+$ type.}
\end{figure}

The F$^+$/F$^-$ system is not the most common form of bacterial conjugation.  It is what is known as a {\it narrow spectrum} conjugation mechanism (Tenover 2006), since the F$^-$ plasmid may only be transferred between cells that are from similar strains.  However, it is known that the genes for resistance to various antibiotic drugs have been transferred between distinct strains of bacteria, suggesting that a {\it broad spectrum} conjugation mechanism is likely the important form of HGT leading to the spread of antibiotic drug resistance in bacterial populations (Tenover 2006).  Nevertheless, because all of the bacterial conjugation mechanisms follow a pathway that is similar to the F$^+$/F$^-$ pathway, we will use the F$^+$/F$^-$ system as the basis for developing our mathematical models of conjugation-mediated HGT.

\section{Materials and Methods}

We assume an asexually reproducing bacterial population, where the genome of each bacterium consists of two double-stranded, semiconservatively replicating DNA molecules.  The first DNA molecule contains all of the genes necessary for the proper growth and reproduction of the bacterium itself.  This DNA molecule corresponds to the large, circular chromosome that defines the bacterial genome.  We assume that there exists a wild-type genome characterized by a ``master" DNA sequence.  It is assumed that a bacterium with the master genome has a wild-type fitness, or first-order growth rate constant, given by $ 1 $.  Such a bacterium is termed {\it viable}.  Furthermore, making what is known as the {\it single-fitness-peak} approximation (Tannenbaum and Shakhnovich 2005), we assume that any mutation to the bacterial genome renders the genome defective, so that the bacterium then has a fitness of $ 0 $.  Bacteria with defective genomes are termed {\it unviable}.

The second DNA molecule is the F-plasmid, which we assume consists of two regions.  The first region comprises the various genes necessary for bacterial conjugation itself, i.e. for allowing the plasmid to move between bacteria.  The second region is assumed to encode for the various enzymes conferring resistance to a given antibiotic.  For this initial study, we are interested in the interplay between conjugation-mediated HGT and antibiotic drug resistance at mutation-selection balance (we will consider adaptive dynamics later), and so this is the simplest model that incorporates these various effects.  

As with the single-fitness-peak approximation made for the bacterial genome, for the F-plasmid we assume that there are master sequences for both the conjugation and antibiotic drug resistance regions.  If the region coding for bacterial conjugation corresponds to a given master sequence, then, assuming that the bacterium is also viable, the F-plasmid may move into another viable F$^-$ bacterium.  Otherwise, we assume that plasmid cannot move into another bacterium, in which case the bacterium is treated as an F$^-$ bacterium.  

Similarly, if the region coding for antibiotic drug resistance corresponds to a given master sequence, then we assume that the bacterium is resistant to the antibiotic.  Otherwise, the bacterium is not resistant to the antibiotic, and is assumed to die according to a first-order rate constant given by $ \kappa_D $.  
We assume that only viable bacteria interact with the antibiotic, since non-viable bacteria do not grow and so may be treated as dead.

A given genome may be characterized by a three symbol sequence $ \sigma = \pm \pm \pm $, specifying the state of the viability, conjugation, and resistance portions of the genome, respectively.  A ``+" is taken to signify that the given genome region is identical to the corresponding master sequence, and a ``-" is taken to signify that the given genome region differs from the corresponding master sequence.  

To develop the evolutionary dynamics equations governing this population, we let $ n_{\sigma} $ denote the number of organisms in the population with genome $ \sigma $.  We wish to develop expressions for $ d n_{\sigma}/dt $ for the various $ \sigma $.  Since we are only interested in the viable population, the $ \sigma $ of interest are $ +++, ++-, +-+, +-- $.

We must now consider the various aspects of the evolutionary dynamics that affect the expressions for the $ d n_{\sigma}/dt $.  The first aspect of the dynamics that we consider is replication:  During the semiconservative replication of the bacterial genome, the strands of the DNA molecule separate and serve as templates for daughter strand synthesis.  Daughter strand synthesis is not necessarily error-free, so that there is a probability $ p $, denoted the replication fidelity, that a given template strand will produce a daughter genome that is identical to the original parent.  Because our genome consists of three genome regions, we may define three such probabilities, denoted $ p_v $, $ p_c $, and $ p_r $, corresponding to the replication fidelities for the viability, conjugation, and resistance portions of the genome.  For a replication fidelity $ p $, it follows that a template strand derived from a master genome region has a probability $ p $ of forming a daughter genome region that is identical to the parent, and a probability of $ 1 - p $ of forming a mutated daughter.  If we assume that sequence lengths are long, then making an assumption known as the {\it neglect of backmutations} (Tannenbaum and Shakhnovich 2005), we assume that a template strand derived from a parent that differs from the master genome produces a daughter that differs from the master genome with probability $ 1 $.  The basis for this assumption is that for very long genomes, mutations will typically occur in previously unmutated regions of the genome, so that mutations will tend to accumulate.

The second aspect of the dynamics that we consider is conjugation:  We assume that conjugation occurs between a viable F$^+$-bacterium and a viable F$^-$-bacterium.  Thus, conjugation can only occur between a bacterium of type $ ++\pm $ and a bacterium of type $ +-\pm $.  This process is modeled as a second-order collision reaction with a rate constant $ \gamma $.  The conjugation process itself involves the transfer of one of the strands of the plasmid from the F$^+$-bacterium to the F$^-$-bacterium, so that the full plasmid needs to be re-synthesized in both bacteria via daughter strand synthesis.  This introduces the possibility of replication errors in either one of the bacteria.

It should be emphasized that we are assuming for simplicity that all bacteria in the population contain exactly one plasmid.  This plasmid may contain the correct copies of the genes for conjugation, in which case the bacterium is an F$^+$-bacterium, or the plasmid may contain defective copies of the genes for conjugation, in which case the bacterium is an F$^-$-bacterium.  We also assume that, during conjugation, the plasmid transferred from the F$^+$-bacterium replaces the plasmid in the F$^-$-bacterium.  This is a simplifying assumption that will obviously have to be re-examined in future research, where we anticipate developing more accurate models that allow for variable plasmid numbers in the bacterial cell.

Putting everything together, we obtain that the evolutionary dynamics equations are,
\begin{widetext}
\begin{eqnarray}
&   &
\frac{d n_{+++}}{dt} = [2 p_v p_c p_r - 1 + \frac{\gamma}{V} (2 p_c p_r - 1) (n_{+-+} + n_{+--})] n_{+++}
\nonumber \\
&   &
\frac{d n_{++-}}{dt} = [2 p_v p_c - 1 - \kappa_D + \frac{\gamma}{V} (2 p_c - 1) (n_{+-+} + n_{+--})] n_{++-} 
\nonumber \\
&   &
+ 2 p_c (1 - p_r) [p_v + \frac{\gamma}{V} (n_{+-+} + n_{+--})] n_{+++}
\nonumber \\
&   &
\frac{d n_{+-+}}{dt} = [2 p_v p_r - 1 - \frac{\gamma}{V} (n_{+++} + n_{++-})] n_{+-+} 
+ 2 (1 - p_c) p_r [p_v + \frac{\gamma}{V} (n_{+-+} + n_{+--})] n_{+++}
\nonumber \\
&   &
\frac{d n_{+--}}{dt} = [2 p_v - 1 - \kappa_D - \frac{\gamma}{V} (n_{+++} + n_{++-})] n_{+--} 
+ 2 (1 - p_c) (1 - p_r) [p_v + \frac{\gamma}{V} (n_{+-+} + n_{+--})] n_{+++} 
\nonumber \\
&   &
+ 2 (1 - p_c) [p_v + \frac{\gamma}{V} (n_{+-+} + n_{+--})] n_{++-} + 2 p_v (1 - p_r) n_{+-+} 
\end{eqnarray}
\end{widetext}
where $ V $ is defined as the system volume.  To put the equations into a form that makes the analysis of the mutation-selection balance possible, we define the total population $ n = n_{+++} + n_{++-} + n_{+-+} + n_{+--} + n_{-++} + n_{-+-} + n_{--+} + n_{---} $, and then define population fractions $ x_{\sigma} $ via $ x_{\sigma} = n_{\sigma}/n $.  We also define a population density $ \rho = n/V $, and we assume that $ \rho $ is constant.  Converting from population numbers to population fractions, we obtain,
\begin{widetext}
\begin{eqnarray}
&   &
\frac{d x_{+++}}{dt} = [2 p_v p_c p_r - 1 + \gamma \rho (2 p_c p_r - 1) (x_{+-+} + x_{+--}) - \bar{\kappa}(t)] x_{+++}
\nonumber \\
&   &
\frac{d x_{++-}}{dt} = [2 p_v p_c - 1 - \kappa_D + \gamma \rho (2 p_c - 1) (x_{+-+} + x_{+--}) - \bar{\kappa}(t)] x_{++-} 
\nonumber \\
&   &
+ 2 p_c (1 - p_r) [p_v + \gamma \rho (x_{+-+} + x_{+--})] x_{+++}
\nonumber \\
&   &
\frac{d x_{+-+}}{dt} = [2 p_v p_r - 1 - \gamma \rho (x_{+++} + x_{++-}) - \bar{\kappa}(t)] x_{+-+} 
+ 2 (1 - p_c) p_r [p_v + \gamma \rho (x_{+-+} + x_{+--})] x_{+++}
\nonumber \\
&   &
\frac{d x_{+--}}{dt} = [2 p_v - 1 - \kappa_D - \gamma \rho (x_{+++} + x_{++-}) - \bar{\kappa}(t)] x_{+--} 
\nonumber \\
&   &
+ 2 (1 - p_c) (1 - p_r) [p_v + \gamma \rho (x_{+-+} + x_{+--})] x_{+++} 
\nonumber \\
&   &
+ 2 (1 - p_c) [p_v + \gamma \rho (x_{+-+} + x_{+--})] x_{++-} + 2 p_v (1 - p_r) x_{+-+} 
\end{eqnarray}
\end{widetext}
where $ \bar{\kappa}(t) = (1/n) (dn/dt) = x_{+++} + x_{+-+} + (1 - \kappa_D) (x_{++-} + x_{+--}) $ is the mean fitness of the population.  In the subsequent analysis, we will be interested in computing the mean fitness at mutation-selection balance, since the mean fitness provides the measure of the effective first-order growth constant of the population.  Therefore, the mean fitness will allow us to understand the selective advantage of HGT in a static environment.

To determine the values for $ p_v $, $ p_c $, and $ p_r $, we assume that daughter strand synthesis has a per-base mismatch probability $ \epsilon $, which incorporates all DNA error-correction mechanisms such as proofreading and mismatch repair.  Because we are assuming complementary double-stranded DNA molecules, we assume that all post-replication mismatches are corrected via various lesion repair mechanisms (e.g. Nucleotide Excision Repair or NER).  However, because at this stage there is no discrimination between parent and daughter strands, a mismatch is either correctly repaired with probability $ 1/2 $, or is fixed as a mutation in the genome with probability $ 1/2 $.  Thus, the net per-base mismatch probability is $ \epsilon/2 $.  If the total sequence length is $ L $, then the probability of producing a mutation-free daughter from a given parent template strand is $ (1 - \epsilon/2)^L $.

If we define $ \mu = L \epsilon $, so that $ \mu $ is the average number of mismatches per template strand per replication cycle, and if we assume that $ L \rightarrow \infty $ while $ \mu $ is held constant, then we obtain that $ (1 - \epsilon/2)^L \rightarrow e^{-\mu/2} $.  For the case of the three-gene model we are considering, we let $ L_v $, $ L_c $, and $ L_r $ denote the lengths of the genome controlling viability, conjugation, and resistance, respectively.  Defining $ L = L_v + L_c + L_r $, and $ \alpha_v = L_v/L $, $ \alpha_c = L_c/L $, $ \alpha_r = L_r/L $, we then obtain that,
\begin{eqnarray}
&   &
p_v = e^{-\alpha_v \mu/2}
\nonumber \\
&   &
p_c = e^{-\alpha_c \mu/2}
\nonumber \\
&   &
p_r = e^{-\alpha_r \mu/2}
\end{eqnarray}

It should be noted that holding $ \mu $ constant in the limit of infinite genome length is equivalent to assuming a fixed per genome replication fidelity in the limit of long genomes.

\section{Results and Discussion}

In this section, we will solve for the mean fitness at mutation-selection balance, denoted by $ \bar{\kappa} $, for two different sets of parameter regimes:  We will first consider the case of arbitrary $ \kappa_D $, but with $ \gamma \rho \rightarrow 0 $ and $ \gamma \rho \rightarrow \infty $.  We will then consider the case of arbitrary $ \gamma \rho $, but with $ \kappa_D \rightarrow 0 $ and $ \kappa_D \rightarrow \infty $.  Both sets of cases are analytically solvable, and may be used to qualitatively understand the behavior of $ \bar{\kappa} $ for arbitrary values of $ \kappa_D $ and $ \gamma \rho $.

In order to avoid having the derivation of the results interfere with the results themselves, for convenience we present the final analytical results for each parameter regime being considered, and then provide the derivations in a subsequent subsection.  We do not relegate the derivations to an appendix, as we believe that they are sufficiently interesting to remain part of the main text. 

\subsection{Behavior of $ \bar{\kappa} $ for arbitrary $ \kappa_D $}

In the limit where $ \gamma \rho \rightarrow 0 $, the ability for conjugation is lost due to genetic drift (since it is never used), and we obtain that,
\begin{equation}
\bar{\kappa}_{\gamma \rho \rightarrow 0} = \max\{2 p_v p_r - 1, 2 p_v - 1 - \kappa_D\}
\end{equation}

We now consider the limit where $ \gamma \rho \rightarrow \infty $.  We obtain at steady-state that,
\begin{equation}
\bar{\kappa}_{\gamma \rho \rightarrow \infty} = \max\{\frac{2 p_v p_c p_r - 1 + 2 (1 - p_v) (1 - p_c)}{2 p_c - 1}, 2 p_v - 1 - \kappa_D\}
\end{equation}
where $ x_{+++} > 0 $ when $ \bar{\kappa} $ is given by the first expression, and $ x_{+++} = 0 $ when $ \bar{\kappa} $ is given by the second expression.

We can also show that $ \bar{\kappa}_{\gamma \rho \rightarrow \infty} < \bar{\kappa}_{\gamma \rho \rightarrow 0} $.

\subsection{Behavior of $ \bar{\kappa} $ for arbitrary $ \gamma \rho $}

Now we consider the behavior of $ \bar{\kappa} $ for arbitrary values of $ \gamma \rho $, but where $ \kappa_D $ is either very small or very large.  Combined with the results of the previous subsection, we may then piece together a qualitative sketch of how $ \bar{\kappa} $ depends on $ \kappa_D $ and $ \gamma \rho $. 

When $ \kappa_D \rightarrow 0 $, there is no selective advantage for maintaining antibiotic drug resistance genes in the genome, and so we expect these genes to be lost to genetic drift.  Thus, we expect, at mutation-selection balance, that $ x_{+++} = x_{+-+} = 0 $, so we need only consider the populations $ x_{++-} $ and $ x_{+--} $.  We may also show that $ \bar{\kappa} = 2 p_v - 1 $.

Furthermore, the fraction of viable conjugators, $ x_{+++} + x_{++-} $, exhibits a transition as a function of $ \gamma \rho $.  For sufficiently small values of $ \gamma \rho $, we have that $ x_{+++} + x_{++-} = 0 $, while for sufficiently large values of $ \gamma \rho $, we have that,
\begin{equation}
x_{+++} + x_{++-} = 2 p_v - 1 - \frac{2 p_v (1 - p_c)}{\gamma \rho (2 p_c - 1)}
\end{equation}

The transition between the two regimes may be shown to occur at,
\begin{equation}
(\gamma \rho)_{trans} \equiv \frac{2 p_v (1 - p_c)}{(2 p_v - 1) (2 p_c - 1)}
\end{equation}

It may be shown that the disappearance of the conjugators below the critical value of $ \gamma \rho $ corresponds to a localization to delocalization transition over the portion of the plasmid coding for conjugation, so that this transition is a conjugation-mediated HGT analogue of the well-known error catastrophe from quasispecies theory (Tannenbaum and Shakhnovich 2005).

To understand this behavior, we note that plasmids with defective genes for conjugation nevertheless replicate due to the replication of the bacteria in which they reside.  Thus, for plasmids with functional genes for conjugation to be preserved in the population, their additional growth rate due to conjugation must overcome the loss of functionality due to replication mistakes in the genes controlling conjugation.  If the conjugation rate is too slow and unable to overcome this loss of functionality, then the fraction of conjugators in the population drops to zero.  

We now consider the case where $ \kappa_D \rightarrow \infty $.  In contrast to the case where $ \gamma \rho \rightarrow \infty $ of the previous subsection, where we could solve for $ \bar{\kappa} $ for arbitrary values of $ \kappa_D $, here we cannot readily analytically solve for $ \bar{\kappa} $ for arbitrary values of $ \gamma \rho $.  However, we can obtain analytical solutions for $ \bar{\kappa} $ in certain limiting cases of $ \gamma \rho $, and then interpolate between the two solution regimes.  As will be seen in the subsection comparing theory and simulation, this approach turns out to be fairly accurate.

In the first limiting case, we assume that $ \gamma \rho $ remains finite in the limit that $ \kappa_D \rightarrow \infty $.  This assures that $ x_{++-} = x_{+--} = 0 $, since the rate of death due to the presence of antibiotics is so fast that no non-resistant genotypes are present in the population.  The fact that $ \gamma \rho $ is taken to be finite in the limit that $ \kappa_D \rightarrow \infty $ means that a non-resistant genotype cannot be ``rescued" via conjugation with a resistant bacterium before death occurs.    

We then obtain that either $ \bar{\kappa} = 2 p_v p_r - 1 $ , or that $ \bar{\kappa} $ is the solution to the following equation:
\begin{widetext}
\begin{equation}
\gamma \rho = \frac{2 (1 - p_r)}{2 p_c p_r - 1} \frac{\bar{\kappa} + 2 (1 - p_v)}{\bar{\kappa}}
 \frac{(\bar{\kappa} + 1 - 2 p_v p_c p_r)^2}{[1 - 2 p_r (1 - p_c)] \bar{\kappa} - [2 p_v p_c p_r - 1 + 2 p_r (1 - p_v) (1 - p_c)]}
\end{equation}
\end{widetext}

In the first case, we have that $ x_{+++} = 0 $, while in the second case we have that $ x_{+++} > 0 $.  The transition between the two regimes may be shown to occur at,
\begin{equation}
(\gamma \rho)_{trans} = \frac{2 p_v p_r (1 - p_c) [1 - 2 p_v (1 - p_r)]}{(2 p_v p_r - 1) (2 p_c p_r - 1)}
\end{equation}
where $ x_{+++} = 0 $ for $ \gamma \rho \leq (\gamma \rho)_{trans} $ and $ x_{+++} > 0 $ for $ \gamma \rho > (\gamma \rho)_{trans} $.  We may show that this expression for $ (\gamma \rho)_{trans} $ is larger than the corresponding expression for the $ \kappa_D = 0 $ case.

To understand the behavior of $ \bar{\kappa} $ where $ \gamma \rho > (\gamma \rho)_{trans} $, we consider the asymptotic behavior of $ \bar{\kappa} $ in the limit as $ \gamma \rho \rightarrow \infty $.  In this case, Eq. (8) reduces to,
\begin{equation}
\bar{\kappa} = \frac{2 p_v p_c p_r -1 + 2 p_r (1 - p_v) (1 - p_c)}{1 - 2 p_r (1 - p_c)}
\end{equation}
We may show that this expression is smaller than the expression for $ \bar{\kappa} $ obtained in the arbitrary $ \kappa_D $, infinite $ \gamma \rho $ case.

We now consider the second limiting case in the $ \kappa_D \rightarrow \infty $ limit, specifically where $ \gamma \rho $ is itself infinite.  Here, however, the ratio between $ \kappa_D $ and $ \gamma \rho $ may play an important role in the competition between death of non-resistant bacteria, and their ``rescue" by conjugation with resistant bacteria.  Thus, here, we will assume that both $ \gamma \rho, \kappa_D \rightarrow \infty $, but we will take $ \gamma \rho/\kappa_D $ to have some given value in this limit.  For large values of this ratio, we expect the rescue effect to dominate over bacterial death, and so the value of $ \bar{\kappa} $ should approach the value obtained for arbitrary $ \kappa_D $ in the $ \gamma \rho \rightarrow \infty $ limit.  For small values of this ratio, we expect bacterial death to dominate over conjugation, and so the value of $ \bar{\kappa} $ should decrease to a value that will need to be determined.

We may show that,
\begin{widetext}
\begin{equation}
\frac{\gamma \rho}{\kappa_D} = \frac{\bar{\kappa} + 2 (1 - p_v)}{\bar{\kappa}}
\frac{[1 - 2 p_r (1 - p_c)] \bar{\kappa} - [2 p_v p_c p_r - 1 + 2 p_r (1 - p_v) (1 - p_c)]}
{[2 p_v p_c p_r - 1 + 2 (1 - p_v) (1 - p_c)] - (2 p_c - 1) \bar{\kappa}}
\end{equation}
\end{widetext}
and so obtain that,
\begin{eqnarray}
&   &
\bar{\kappa}_{\gamma \rho/\kappa_D \rightarrow 0} = \frac{2 p_v p_c p_r - 1 + 2 p_r (1 - p_v) (1 - p_c)}{1 - 2 p_r (1 - p_c)}
\nonumber \\
&   &
\bar{\kappa}_{\gamma \rho/\kappa_D \rightarrow \infty} = \frac{2 p_v p_c p_r - 1 + 2 (1 - p_v)(1 - p_c)}{2 p_c - 1}
\end{eqnarray}

Therefore, for large $ \kappa_D $, we expect that $ \bar{\kappa} $ will initially be given by $ 2 p_v p_r - 1 $ up to a critical value of $ \gamma \rho $, after which it begins to decrease according to Eq. (8).  Once $ \gamma \rho $ becomes sufficiently large, we expect that the $ \gamma \rho/\kappa_D $ ratio is such that the functional form for $ \bar{\kappa} $ transitions from the finite $ \gamma \rho $ solution to the infinite $ \gamma \rho $, fixed $ \gamma \rho/\kappa_D $ solution.  To estimate the transition point between the two solution regimes, we equate the values for $ \gamma \rho $ as a function of $ \bar{\kappa} $ for the two solutions.  This allows us to solve for $ \bar{\kappa} $ and thereby allow us to solve for $ \gamma \rho $.  

We then obtain that the transition point occurs at,
\begin{widetext}
\begin{equation}
(\frac{\gamma \rho}{\sqrt{\kappa_D}})_{trans} = 2 p_r \frac{2 p_c p_r - 1 + 2 (1 - p_v) (1 - p_r)}{2 p_v p_c p_r - 1 + 2 p_r (1 - p_v) (1 - p_c)}
\sqrt{\frac{p_v (1 - p_c)}{1 - 2 p_r (1 - p_c)}}
\end{equation}
\end{widetext}

Note that, as $ \kappa_D \rightarrow \infty $, we have that $ (\gamma \rho)_{trans} \rightarrow \infty $ and $ (\gamma \rho/\kappa_D)_{trans} \rightarrow 0 $, so the assumptions that allowed us to make the calculation above are valid.

\subsection{Comparison of Theory and Simulation}

Figure 2 shows plots of $ \bar{\kappa} $ versus $ \mu $ for both the $ \gamma \rho \rightarrow 0 $, $ \gamma \rho \rightarrow \infty $ limits.  Plots were obtained using both the analytical formulas obtained in this paper, as well as via stochastic simulations of replicating organisms.  Note the good agreement between theory and simulation.  

\begin{figure}
\includegraphics[width = 0.7\linewidth, angle = -90]{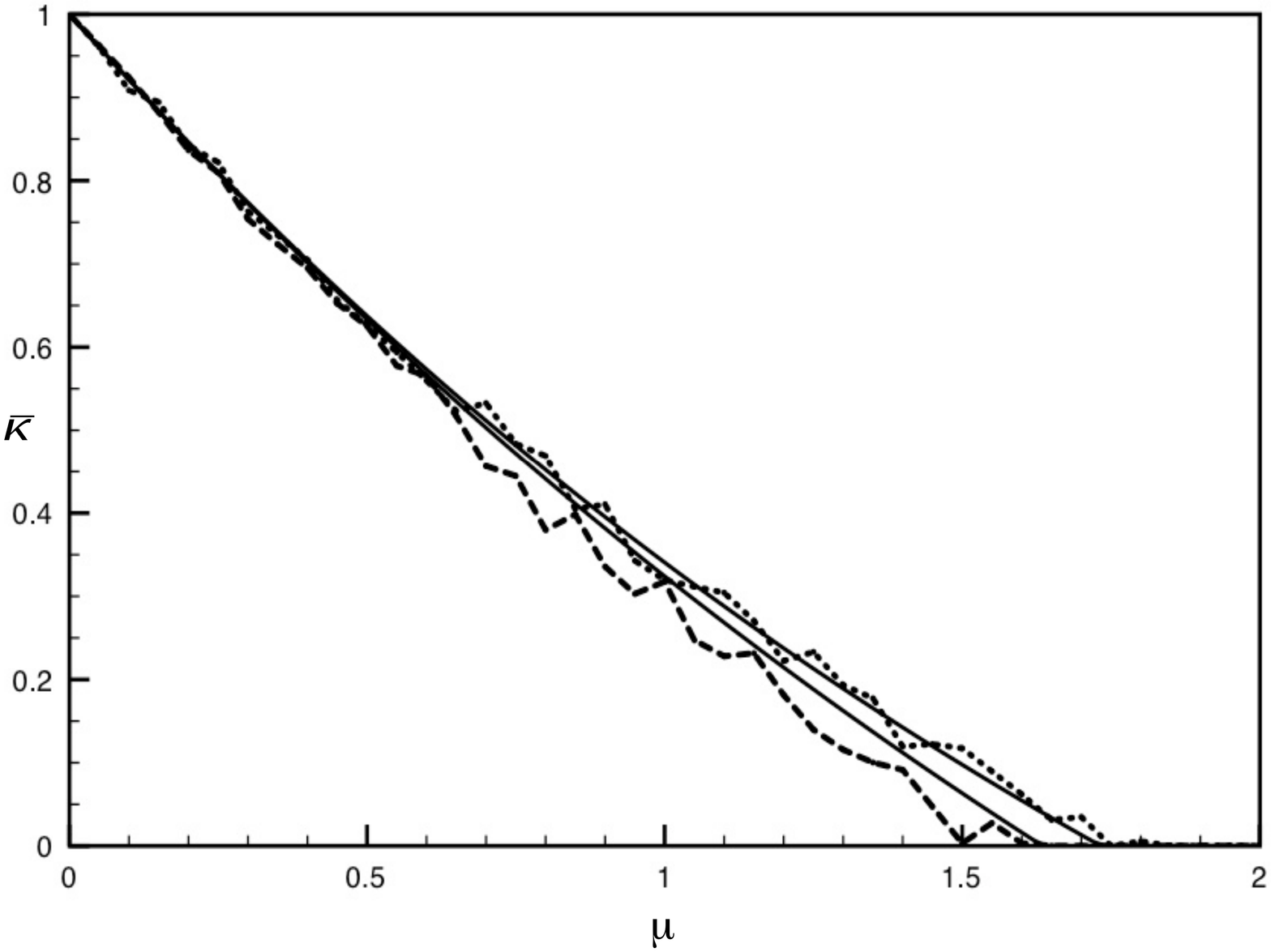}
\caption{Plots of $ \bar{\kappa} $ versus $ \mu $ for both the $ \gamma \rho \rightarrow 0 $, $ \gamma \rho \rightarrow \infty $ limits.  The parameter values we took are $ \alpha_v = 0.6 $, $ \alpha_c = \alpha_r = 0.2 $, and $ \kappa_D = 10 $.  We show both analytical results and results from stochastic simulations.  The analytical results are plotted using thin solid lines, where the top curve corresponds to the $ \gamma \rho = 0 $ result, while the bottom curve corresponds to the $ \gamma \rho = \infty $ result.  The dotted line corresponds to the stochastic simulation for $ \gamma \rho = 0 $, and the dashed line corresponds to the stochastic simulation for $ \gamma \rho = \infty $.  Parameter values for the stochastic simulations were $ L_v = 30 $, $ L_c = L_r = 10 $, and a population size of $ 1,000 $.}
\end{figure}

Figure 3 illustrates the regimes, as a function of $ \mu $ and $ \gamma \rho $, where there exist a positive fraction of conjugators at steady-state, and where the fraction of conjugators is zero.  This is computed for the $ \kappa_D = 0 $ limit.  Note that, as $ \mu $ increases, $ \gamma \rho $ must be pushed to higher values so that there is a positive fraction of conjugators at steady-state.  As explained before, this increase in $ \gamma \rho $ is necessary to overcome the mutation-induced loss of functionality as $ \mu $ increases.

\begin{figure}
\includegraphics[width = 0.7\linewidth, angle = -90]{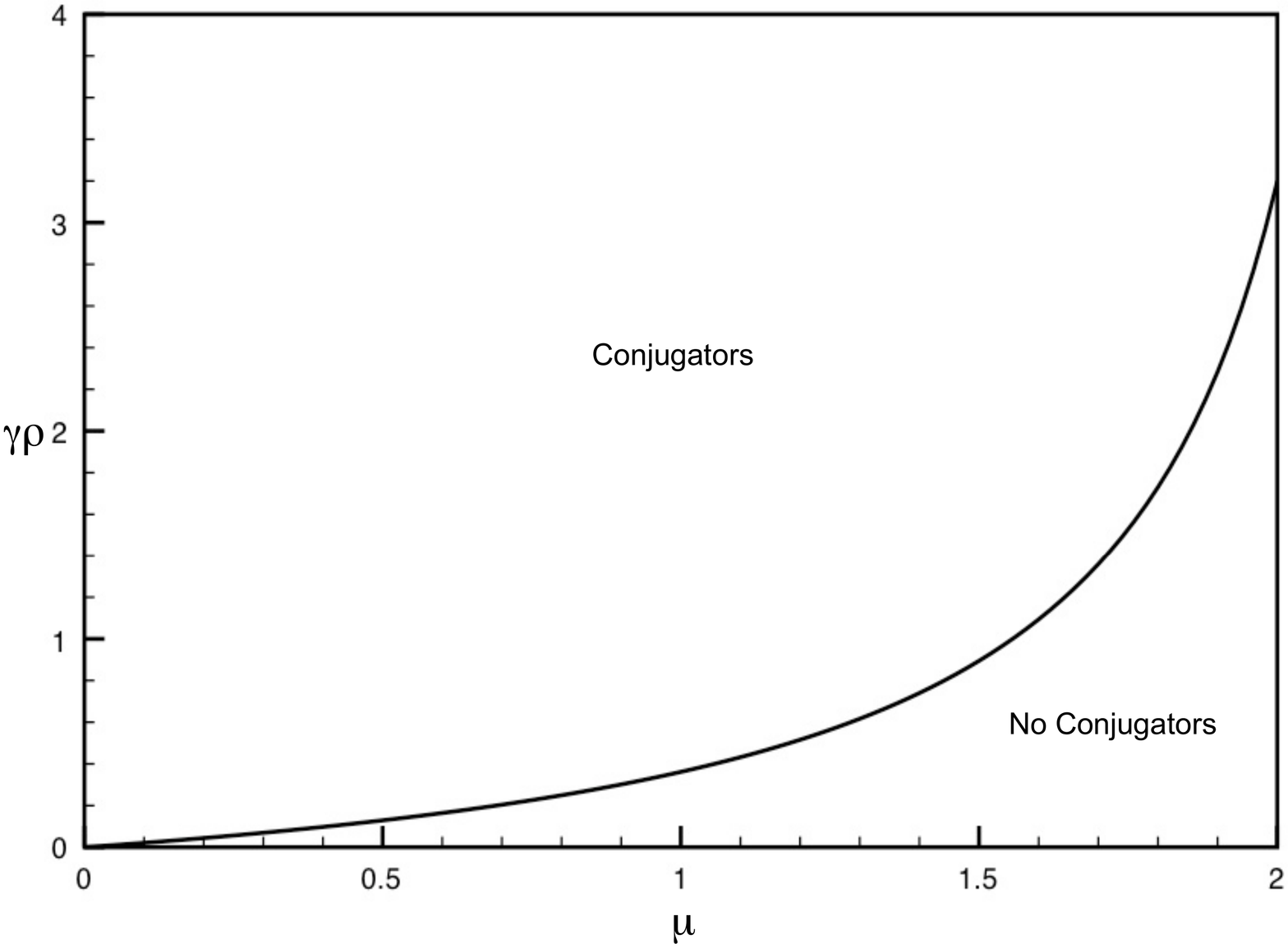}
\caption{Regimes of existence and non-existence of conjugators as a function of $ \mu $ and $ \gamma \rho $, where $ \kappa_D = 0 $.  The boundary between the two regimes was computed analytically.}
\end{figure}

Figure 4 shows three plots of $ \bar{\kappa} $ versus $ \gamma \rho $ for $ \kappa_D = 10 $.  One of the plots was obtained by numerically solving for the mutation-selection balance using fixed-point iteration.  The other two plots correspond to the infinite $ \kappa_D $, finite $ \gamma \rho $, and infinite $ \kappa_D $, fixed $ \gamma \rho/\kappa_D $ expressions for $ \bar{\kappa} $ given in the preceding subsections.  Note that already for $ \kappa_D = 10 $ the approximate analytical solutions capture the dependence of $ \bar{\kappa} $ on $ \gamma \rho $ fairly accurately.

\begin{figure}
\includegraphics[width = 0.7\linewidth, angle = -90]{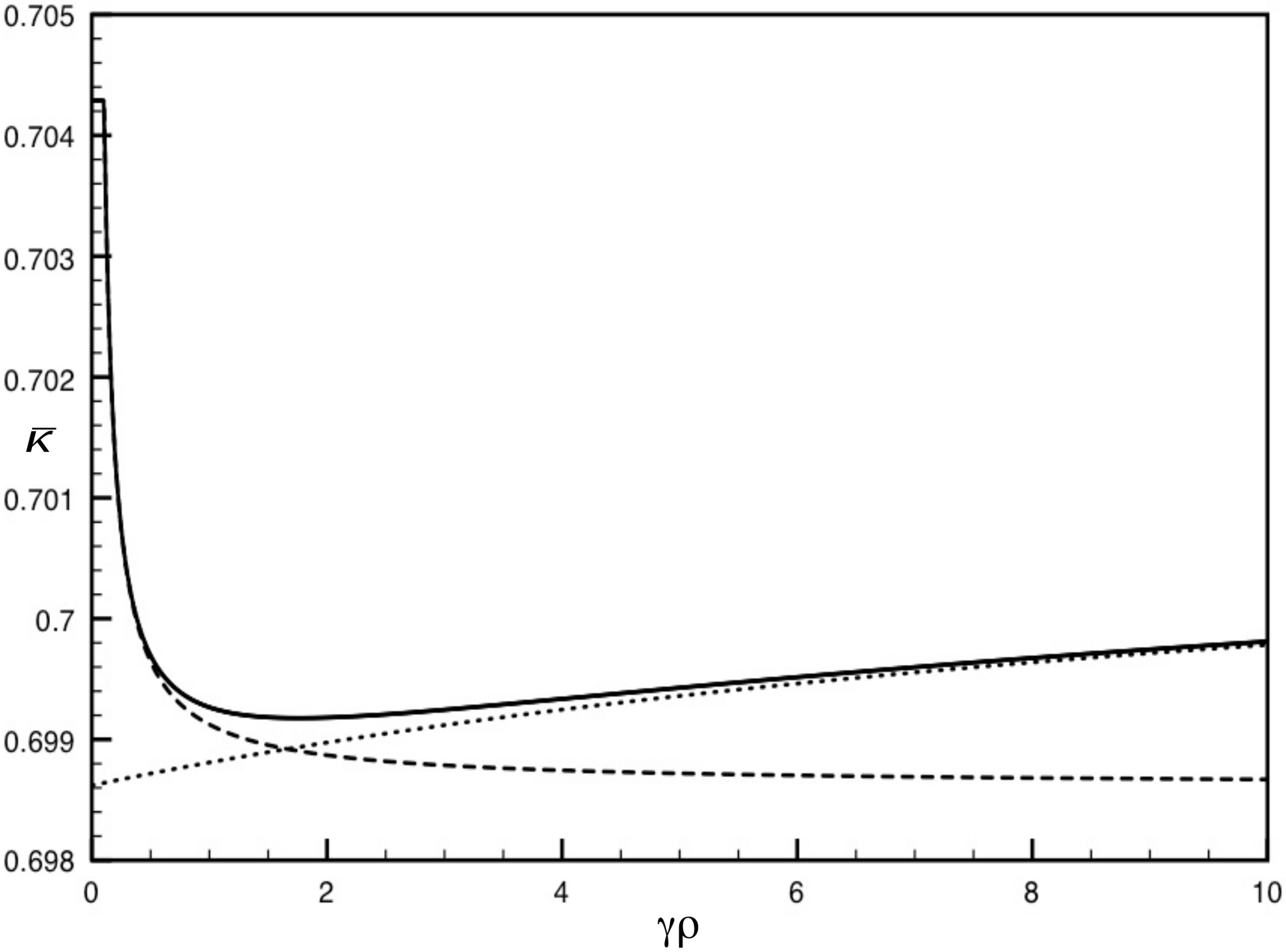}
\caption{Plots of $ \bar{\kappa} $ versus $ \gamma \rho $ for $ \kappa_D = 10 $, $ \mu = 0.4 $, $ \alpha_v = 0.6 $, $ \alpha_c = \alpha_r = 0.2 $.  The plot marked with the solid line was obtained by numerically solving for $ \bar{\kappa} $ using fixed-point iteration.  The dashed line was obtained by using the infinite $ \kappa_D $, finite $ \gamma \rho $ expression for $ \bar{\kappa} $, while the dotted line was obtained by using the infinite $ \kappa_D $, fixed $ \gamma \rho/\kappa_D $ expression for $ \bar{\kappa} $.}
\end{figure}

\subsection{Derivation Details of the Analytical Results}

\subsubsection{Derivation of $ \bar{\kappa} $ for arbitrary $ \kappa_D $, and $ \gamma \rho \rightarrow 0 $}

Due to the nature of exponential growth, for the population fractions to converge to a stable steady-state we must have that, $ \bar{\kappa} \geq 2 p_v p_c p_r - 1, 2 p_v p_c - 1 - \kappa_D, 2 p_v p_r - 1, 2 p_v - 1 - \kappa_D $.  Because $ 2 p_v p_c p_r - 1 < 2 p_v p_r - 1 $, and $ 2 p_v p_c - 1 - \kappa_D < 2 p_v - 1 - \kappa_D $, it follows that $ \bar{\kappa} \geq 2 p_v p_r - 1, 2 p_v - 1 - \kappa_D $.  However, if we then look at the steady-state version of Eq. (2), obtained by setting the time derivatives to $ 0 $, we then obtain that $ x_{+++} = x_{++-} = 0 $.  If $ x_{+-+} > 0 $, then the third equation gives us that $ \bar{\kappa} = 2 p_v p_r - 1 $, otherwise the fourth equation gives us $ \bar{\kappa} = 2 p_v - 1 - \kappa_D $. 

So, we have shown that $ \bar{\kappa} \geq 2 p_v p_r - 1, 2 p_v - 1 - \kappa_D $, and yet $ \bar{\kappa} = 2 p_v p_r - 1 $ or $ 2 p_v - 1 - \kappa_D $.  These two requirements imply that $ \bar{\kappa} = \max\{2 p_v p_r - 1, 2 p_v - 1 - \kappa_D\} $.  Note that we have also shown that $ x_{+++} + x_{++-} = 0 $, so that our claim that conjugation is lost due to genetic drift has also been proven.

\subsubsection{Derivation of $ \bar{\kappa} $ for arbitrary $ \kappa_D $, and $ \gamma \rho \rightarrow \infty $}

In the limit where $ \gamma \rho \rightarrow \infty $, we have that $ x_{+-+} = x_{+--} = 0 $.  However, $ \gamma \rho x_{+-+} $ and $ \gamma \rho x_{+--} $ may converge to positive values.  So, we define $ z_{+-+} = \gamma \rho x_{+-+} $ and $ z_{+--} = \gamma \rho x_{+--} $.

Because $ x_{+-+} = x_{+--} = 0 $, we also have that $ d x_{+-+}/dt = d x_{+--}/dt = 0 $, and so from Eq. (2) we have that,
\begin{eqnarray}
&   &
0 = -z_{+-+} (x_{+++} + x_{++-}) 
\nonumber \\
&   &
+ 2 (1 - p_c) [p_v + z_{+-+} + z_{+--}] p_r x_{+++}
\nonumber \\
&   &
0 = -z_{+--} (x_{+++} + x_{++-}) 
\nonumber \\
&   &
+ 2 (1 - p_c) [p_v + z_{+-+} + z_{+--}] [(1 - p_r) x_{+++} + x_{++-}]
\nonumber \\
\end{eqnarray}

Summing these two equations and solving for $ z_{+-+} + z_{+--} $ gives,
\begin{equation}
z_{+-+} + z_{+--} = \frac{2 (1 - p_c) p_v}{2 p_c - 1}
\end{equation}

Substituting into the expressions for $ d x_{+++}/dt $ and $ d x_{++-}/dt $ from Eq. (2) we obtain, after some manipulation,
\begin{eqnarray}
&   &
\frac{d x_{+++}}{dt} = [\frac{2 p_v p_c p_r - 1 + 2 (1 - p_v)(1 - p_c)}{2 p_c - 1} - \bar{\kappa}(t)] x_{+++}
\nonumber \\
&   &
\frac{d x_{++-}}{dt} = [2 p_v - 1 - \kappa_D - \bar{\kappa}(t)] x_{++-} + \frac{2 p_v p_c (1 - p_r)}{2 p_c - 1} x_{+++}
\nonumber \\
\end{eqnarray}

Following a similar argument to the $ \gamma \rho \rightarrow 0 $ case, we obtain the expression for $ \bar{\kappa}_{\gamma \rho \rightarrow \infty} $ given above.

To prove that $ \bar{\kappa}_{\gamma \rho \rightarrow \infty} < \bar{\kappa}_{\gamma \rho \rightarrow 0} $, we need only show that,
\begin{equation}
\frac{2 p_v p_c p_r - 1 + 2 (1 - p_v)(1 - p_c)}{2 p_c - 1} < 2 p_v p_r - 1
\end{equation}
After some manipulation, it may be shown that this inequality is equivalent to, $ p_r < 1 $, which clearly holds, thereby proving the claim.

\subsubsection{Derivation of $ \bar{\kappa} $ for $ \kappa_D \rightarrow 0 $, and arbitrary $ \gamma \rho $}

We can add the first two equations from Eq. (2), and also the third and fourth equations, to obtain the pair of equations,
\begin{widetext}
\begin{eqnarray}
&   &
\frac{d (x_{+++} + x_{++-})}{dt} = [2 p_v p_c - 1 + \gamma \rho (2 p_c - 1) (x_{+-+} + x_{+--}) - \bar{\kappa}(t)] (x_{+++} + x_{++-}) 
\nonumber \\
&   &
\frac{d (x_{+-+} + x_{+--})}{dt} = [2 p_v - 1 - \gamma \rho (x_{+++} + x_{++-}) - \bar{\kappa}(t)] (x_{+-+} + x_{+--})
\nonumber \\
&   &
+ 2 (1 - p_c) [p_v + \gamma \rho (x_{+-+} + x_{+--})] (x_{+++} + x_{++-})
\end{eqnarray}
\end{widetext}
Summing these two equations then gives,
\begin{widetext}
\begin{equation}
\frac{d (x_{+++} + x_{++-} + x_{+-+} + x_{+--})}{dt} = [2 p_v - 1 - \bar{\kappa}(t)] (x_{+++} + x_{++-} + x_{+-+} + x_{+--}) 
\end{equation}
\end{widetext}
from which it follows that $ \bar{\kappa} = 2 p_v - 1 $ at steady-state.

Substituting this value for $ \bar{\kappa} $ into the steady-state version of Eq. (18), we obtain,
\begin{equation}
0 = [(2 p_c - 1) \gamma \rho (x_{+-+} + x_{+--}) - 2 p_v (1 - p_c)] (x_{+++} + x_{++-})
\end{equation}
which gives either that $ x_{+++} + x_{++-} = 0 $ or $ x_{+-+} + x_{+--} = 2 p_v (1 - p_c)/[\gamma \rho (2 p_c - 1)] $.  If the second case holds, then since 
$ 2 p_v - 1 = \bar{\kappa} = x_{+++} + x_{++-} + x_{+-+} + x_{+--} $, we obtain that,
\begin{equation}
x_{+++} + x_{++-} = 2 p_v - 1 - \frac{2 p_v (1 - p_c)}{\gamma \rho (2 p_c - 1)}
\end{equation}
Now, for large values of $ \gamma \rho $, we expect that the population will consist of a non-zero fraction of conjugators, so that $ x_{+++} + x_{++-} > 0 $.  However, because $ x_{+++} + x_{++-} $ cannot be negative, we must have that,
\begin{equation}
\gamma \rho \geq (\gamma \rho)_{trans} \equiv \frac{2 p_v (1 - p_c)}{(2 p_v - 1) (2 p_c - 1)}
\end{equation}
in order for $ x_{+++} + x_{++-} \geq 0 $.  Therefore, by continuity, we expect that $ x_{+++} + x_{++-} = 0 $ for $ \gamma \rho \leq (\gamma \rho)_{trans} $, and $ x_{+++} + x_{++-} = 2 p_v - 1 - \frac{2 p_v (1 - p_c)}{\gamma \rho (2 p_c - 1)} > 0 $ for $ \gamma \rho > (\gamma \rho)_{trans} $.

\subsubsection{Derivation of $ \bar{\kappa} $ for $ \kappa_D \rightarrow \infty $, and finite $ \gamma \rho $}

In this limiting case, although $ x_{++-} = x_{+--} = 0 $, it is possible that $ y_{++-} \equiv \kappa_D x_{++-} $ and $ y_{+--} \equiv \kappa_D x_{+--} $ have non-zero, finite values in the limit as $ \kappa_D \rightarrow \infty $, and so we need to consider the effect of these quantities in our analysis.  We then have that the steady-state version of Eq. (2) reads,
\begin{widetext}
\begin{eqnarray}
&   &
0 = [2 p_v p_c p_r - 1 + \gamma \rho (2 p_c p_r - 1) x_{+-+} - \bar{\kappa}] x_{+++}
\nonumber \\
&   &
0 = [2 p_v p_r - 1 - \gamma \rho x_{+++} - \bar{\kappa}] x_{+-+} + 2 (1 - p_c) p_r [p_v + \gamma \rho x_{+-+}] x_{+++}
\nonumber \\
&   &
y_{++-} = 2 p_c (1 - p_r) [p_v + \gamma \rho x_{+-+}] x_{+++}
\nonumber \\
&   &
y_{+--} = 2 (1 - p_c) (1 - p_r) [p_v + \gamma \rho x_{+-+}] x_{+++} + 2 p_v (1 - p_r) x_{+-+}
\end{eqnarray}
\end{widetext}

If $ x_{+++} = 0 $ at steady-state, then $ \bar{\kappa} = 2 p_v p_r - 1 $.  So, let us consider the case where $ x_{+++} > 0 $.  Summing the first two equations from Eq. (23) gives,
\begin{equation}
2 (1 - p_r) \gamma \rho x_{+++} x_{+-+} = [2 p_v p_r - 1 - \bar{\kappa}] (x_{+++} + x_{+-+})
\end{equation}

Summing the last two equations from Eq. (23) then gives,
\begin{equation}
y_{++-} + y_{+--} = [2 p_v - 1 - \bar{\kappa}] (x_{+++} + x_{+-+})
\end{equation}

Now, in the limiting case being considered here, we have that $ \bar{\kappa} = x_{+++} + x_{+-+} - y_{++-} - y_{+--} = [\bar{\kappa} + 2 (1 - p_v)] (x_{+++} + x_{+-+}) $, and so,
\begin{equation}
x_{+++} + x_{+-+} = \frac{\bar{\kappa}}{\bar{\kappa} + 2 (1 - p_v)}
\end{equation}

Since $ x_{+++} > 0 $, the first equation from Eq. (23) gives,
\begin{equation}
x_{+-+} = \frac{\bar{\kappa} + 1 - 2 p_v p_c p_r}{\gamma \rho (2 p_c p_r - 1)}
\end{equation}
and so,
\begin{equation}
x_{+++} = \frac{\bar{\kappa}}{\bar{\kappa} + 2 (1 - p_v)} - \frac{\bar{\kappa} + 1 - 2 p_v p_c p_r}{\gamma \rho (2 p_c p_r - 1)}
\end{equation}
Substituting into Eq. (24) gives the following non-linear equation that $ \bar{\kappa} $ must satisfy:
\begin{widetext}
\begin{equation}
2 (1 - p_r) \frac{\bar{\kappa} + 1 - 2 p_v p_c p_r}{2 p_c p_r - 1} 
[\frac{\bar{\kappa}}{\bar{\kappa} + 2 (1 - p_v)} - \frac{\bar{\kappa} + 1 - 2 p_v p_c p_r}{\gamma \rho (2 p_c p_r - 1)}]
 = \frac{\bar{\kappa}}{\bar{\kappa} + 2 (1 - p_v)} [2 p_v p_r - 1 - \bar{\kappa}]
\end{equation}
\end{widetext}
which, after some manipulation, may be shown to be equivalent to Eq. (8).

To determine the critical value for the transition between the $ x_{+++} = 0 $ and $ x_{+++} > 0 $ regimes, we note that if $ x_{+++} $ is continuous at this transition, then we must have that $ x_{+++} = 0 $ using the expression in Eq. (28), which gives that $ \bar{\kappa} = 2 p_v p_r - 1 $ from Eq. (29), so that $ \bar{\kappa} $ is also continuous at this transition.  Solving for the critical value of $ \gamma \rho $ then gives,
\begin{equation}
(\gamma \rho)_{trans} = \frac{2 p_v p_r (1 - p_c) [1 - 2 p_v (1 - p_r)]}{(2 p_v p_r - 1) (2 p_c p_r - 1)}
\end{equation}
So, for $ \gamma \rho \leq (\gamma \rho)_{trans} $, we have that $ x_{+++} = 0 $ and $ \bar{\kappa} = 2 p_v p_r - 1 $, while for $ \gamma \rho > (\gamma \rho)_{trans} $ we have that $ x_{+++} > 0 $ and $ \bar{\kappa} $ is given by the solution to Eq. (8) or, equivalently, Eq. (29).

To show that this value for $ (\gamma \rho)_{trans} $ is larger than the corresponding value obtained for $ \kappa_D = 0 $, we need to show that,
\begin{equation}
\frac{2 p_v p_r (1 - p_c) [1 - 2 p_v (1 - p_r)]}{(2 p_v p_r - 1) (2 p_c p_r - 1)} > \frac{2 p_v (1 - p_c)}{(2 p_v - 1) (2 p_c - 1)}
\end{equation}
After some manipulation, this inequality may be shown to be equivalent to,
\begin{equation}
4 p_v p_r (2 p_c - 1) (1 - p_v) + 2 p_v p_r - 1 >  0
\end{equation}
which clearly holds, and so the inequality is established.  

Finally, to show that the value of $ \bar{\kappa} $ as $ \gamma \rho \rightarrow \infty $ is smaller than the value of $ \bar{\kappa} $ obtained in the arbitrary $ \kappa_D $, $ \gamma \rho \rightarrow \infty $ limit, we need to show that,
\begin{eqnarray}
&   &
\frac{2 p_v p_c p_r -1 + 2 p_r (1 - p_v) (1 - p_c)}{1 - 2 p_r (1 - p_c)} 
\nonumber \\
&   &
< \frac{2 p_v p_c p_r - 1 + 2 (1 - p_v) (1 - p_c)}{2 p_c - 1}
\end{eqnarray}
After some manipulation, this condition may be shown to be equivalent to,
\begin{equation}
p_v (2 p_c p_r - 1) (1 - p_c) (1 - p_r) > 0
\end{equation}
which establishes the inequality.

\subsubsection{Derivation of $ \bar{\kappa} $ for $ \kappa_D \rightarrow \infty $, and fixed value of $ \gamma \rho/\kappa_D $}

Because $ \gamma \rho $ is infinite, we expect that $ x_{+-+} = x_{+--} = 0 $, although $ z_{+-+} \equiv \gamma \rho x_{+-+} $ and $ z_{+--} \equiv \gamma \rho x_{+--} $ may converge to positive, though finite, values.  Also, because the $ ++- $ genomes, as conjugators, cannot be ``rescued" by conjugators themselves, we expect that $ x_{++-} = 0 $ in the limit as $ \kappa_D \rightarrow \infty $, though again it is possible that $ y_{++-} \equiv \kappa_D x_{++-} $ converges to a positive value.  We only expect $ x_{+++} > 0 $, since the $ +++ $ genomes are both conjugators and resistant to the antibiotic, and so are not destroyed by conjugation or by antibiotic-induced death.  

The steady-state equations then become,
\begin{widetext}
\begin{eqnarray}
&   &
\bar{\kappa} = 2 p_v p_c p_r - 1 + (2 p_c p_r - 1) (z_{+-+} + z_{+--})
\nonumber \\
&   &
y_{++-} = 2 p_c (1 - p_r) [p_v + z_{+-+} + z_{+--}] x_{+++}
\nonumber \\
&   &
z_{+-+} = 2 (1 - p_c) p_r [p_v + z_{+-+} + z_{+--}]
\nonumber \\
&   &
\frac{\kappa_D}{\gamma \rho} z_{+--} = [2 (1 - p_c) (1 - p_r) (p_v + z_{+-+} + z_{+--}) - z_{+--}] x_{+++} 
\end{eqnarray}
\end{widetext}

From the first equation we have that $ z_{+-+} + z_{+--} = (\bar{\kappa} + 1 - 2 p_v p_c p_r)/(2 p_c p_r - 1) $.  We therefore have that,
\begin{widetext}
\begin{eqnarray}
&   &
y_{++-} = \frac{2 p_c (1 - p_r)}{2 p_c p_r - 1} (\bar{\kappa} + 1 - p_v) x_{+++}
\nonumber \\
&   &
z_{+-+} = \frac{2 (1 - p_c) p_r}{2 p_c p_r - 1} (\bar{\kappa} + 1 - p_v)
\nonumber \\
&   &
z_{+--} = \frac{[1 - 2 p_r (1 - p_c)] \bar{\kappa} - [2 p_v p_c p_r - 1 + 2 p_r (1 - p_v) (1 - p_c)]}{2 p_c p_r - 1}
\nonumber \\
&   &
\frac{\kappa_D}{\gamma \rho} z_{+--} = \frac{2 p_v p_c p_r - 1 + 2 (1 - p_v) (1 - p_c) - (2 p_c - 1) \bar{\kappa}}{2 p_c p_r - 1} x_{+++}
\end{eqnarray}
\end{widetext}
and we also have in this limit that $ \bar{\kappa} = x_{+++} - y_{++-} - \kappa_D/(\gamma \rho) z_{+--} $.  Substituting in the expressions for $ y_{++-} $ and $ \kappa_D/(\gamma \rho) z_{+--} $, we obtain,
\begin{equation}
x_{+++} = \frac{\bar{\kappa}}{\bar{\kappa} + 2 (1 - p_v)}
\end{equation}
Substituting this expression into the last equality of Eq. (36), and using the expression for $ z_{+--} $, gives us Eq. (11).

\subsubsection{Derivation of the transition point between the two functional forms for $ \bar{\kappa} $ for $ \kappa_D \rightarrow \infty $}

Equating the finite $ \gamma \rho $ with the infinite $ \gamma \rho $ expressions for $ \bar{\kappa} $, we obtain that the transition point
occurs where,
\begin{widetext}
\begin{eqnarray}
&   &
[1 - 2 p_r (1 - p_c)] \bar{\kappa} - [2 p_v p_c p_r - 1 + 2 p_r (1 - p_v) (1 - p_c)] 
= \frac{\bar{\kappa} + 1 - 2 p_v p_c p_r}{\sqrt{\kappa_D}} 
\times \nonumber \\
&   &
\sqrt{\frac{2 (1 - p_r)}{2 p_c p_r - 1} ([2 p_v p_c p_r - 1 + 2 (1 - p_v) (1 - p_c)] - (2 p_c - 1) \bar{\kappa})} 
\end{eqnarray}
\end{widetext}

Since $ \kappa_D \rightarrow \infty $, we then obtain that the transition point occurs where the left-hand side is zero, so that $ \bar{\kappa} = [2 p_v p_c p_r - 1 + 2 p_r (1 - p_v) (1 - p_c)]/[1 - 2 p_r (1 - p_c)] $.  To estimate the value of $ \gamma \rho $ where this transition occurs in the limit of large $ \kappa_D $, we substitute 
the expression for $ [1 - 2 p_r (1 - p_c)] \bar{\kappa} - [2 p_v p_c p_r - 1 + 2 p_r (1 - p_v) (1 - p_c)]  $ given in Eq. (38) into Eq. (8), and then substitute the value of $ \bar{\kappa} $ that we obtained for the transition.  After some manipulation, we obtain the expression given by Eq. (13).

\section{Conclusions}

We have developed a mathematical model describing the role that conjugation-mediated Horizontal Gene Transfer (HGT) has on the mutation-selection balance of a unicellular, asexually reproducing, prokaryotic population.  Because HGT is believed to play a major role in the spread of antibiotic drug resistance in bacteria, we considered the effect of an antibiotic on the mutation-selection balance of the population.  Interestingly, we found that, in a static environment at mutation-selection balance, conjugation actually reduces the mean fitness of the population.  However, by studying the dependence of the mean fitness on $ \gamma \rho $ for large values of $ \kappa_D $, the antibiotic-induced first-order death rate constant, we find that the behavior is somewhat more complicated:  For small values of $ \gamma \rho $, the mean fitness is constant, and the fraction of viable conjugators in the population is $ 0 $.  At a critical value of $ \gamma \rho $, the fraction of viable conjugators begins to increase, and the mean fitness decreases to its minimum value.  After reaching its minimum, the mean fitness increases asymptotically to the $ \gamma \rho \rightarrow \infty $ limit, which is nevertheless smaller than the small $ \gamma \rho $ value for the mean fitness.  We developed approximate analytical solutions for the functional dependence of the mean fitness on $ \gamma \rho $ in the limit of large $ \kappa_D $, and found that these solutions agree well with simulation.  It is important to note that the fitness variations as a function of $ \gamma \rho $ were fairly small for the parameter values studied.  Nevertheless, we believe that this is non-trivial behavior that is important to characterize.

Although the results of our paper are based on a highly simplified model, they nevertheless suggest that HGT does not provide a selective advantage in a static environment.  This is likely due to the fact that, due to mutation, HGT can destroy antibiotic drug resistance in a previously resistant cell.  While HGT can also confer resistance to a non-resistant cell, natural selection alone is sufficient to maximize the population mean fitness in a static environment.  HGT simply has the net effect of destroying favorable genes, thereby lowering the mean fitness.  This result may be viewed as an example of the ``If it is not broken, do not fix it" principle.

Thus, based on the results of this paper, we argue that HGT likely only has a selective advantage in dynamic environments, where it would act to speed up rates of adaptation.  While this result needs to be checked in future research, it is nevertheless consistent with the observation that bacteria can regulate their rates of HGT.  For example, it is known that, in response to stress, bacteria can activate the SOS response (Beaber et al. 2004), which has the effect of increasing rates of HGT.  This is consistent with our results suggesting that HGT should be kept at a minimal level in static environments, and increased in dynamic environments.  It is also worth mentioning that while conjugation-mediated HGT has not been specifically modeled before in this manner (at least to our knowledge), other HGT-like models have been studied (Park and Deem 2007; Cohen et al. 2005), and have found that HGT does indeed allow for faster adaptation in dynamic environments (Cohen et al. 2005).       

\bigskip\noindent
{\bf REFERENCES}

\medskip\noindent
Beaber, J.W., Hochhut, B., and Waldor, M.K., 2004 SOS Response Promotes Horizontal Dissemination of Antibiotic Resistance Genes.  Nature (London) 427:  72 - 74.

\medskip\noindent
Cohen, E., Kessler, D.A., and Levine, H., 2005 Recombination Dramatically Speeds Up Evolution of Finite Populations.  Physical Review Letters 94:  098102 (4 pages).

\medskip\noindent
Ochman, H., Lawrence, J.G., and Groisman, E.A., 2000 Lateral Gene Transfer and the Nature of Bacterial Innovation.  Nature (London) 405: 299-304.

\medskip\noindent
Park, J.M., and Deem, M.W., 2007 Phase Diagrams of Quasispecies Theory with Recombination and Horizontal Gene Transfer.  Physical Review Letters 98:  058101 (4 pages).

\medskip\noindent
Russi et al., 2008 Molecular Machinery for DNA Translocation in Bacterial Conjugation.  Plasmids:  Current Research and Future Trends, Caister Academic Press.

\medskip\noindent
Tannenbaum, E., Shakhnovich, E.I., 2005 Semiconservative Replication, Genetic Repair, and Many-Gened Genomes:  Extending the Quasispecies Paradigm to Living Systems.  Physics of Life Reviews 2:  290-317.

\medskip\noindent
Tenover, F., 2006 Mechanisms of Antimicrobial Resistance in Bacteria.  American Journal of Infection Control 34:  S3-S10.

\medskip\noindent
Walsh, C., 2000 Molecular Mechanisms that Confer Antibacterial Drug Resistance.  Nature (London) 406:  775-781.

\end{document}